\documentclass[fp,twocolumn]{jpsj3}
\usepackage{txfonts}
\usepackage{dcolumn}
\usepackage{physics}
\usepackage{bm}% bold math
\usepackage[pdftex]{graphicx}
\usepackage[pdftex]{color}

\begin{document}

\title{Multi-Triplon Excitations of Hubbard Ladders with Site-Dependent Potentials}

\author{
  Nobuya Maeshima$^{1,2}$\thanks{maeshima@ims.tsukuba.ac.jp} and Ken-ichi Hino$^{2,1}$
}

\inst{
  $^1$Center for Computational Sciences,      University of Tsukuba, Tsukuba, Ibaraki 305-8577, Japan \\  
  $^2$Institute of Pure and Applied Sciences, University of Tsukuba, Tsukuba, Ibaraki 305-8573, Japan
}
\recdate{\today}

\abst{
    We study low-lying spin-singlet excitations of two-leg Hubbard ladders with site-dependent potentials.
    Using general formulas of the charge disproportionation induced by the site-dependent potentials, we derive the contributions of spin degrees of freedom to the spectral functions such as the dynamical charge structure factor $N(\bm{k},\omega)$ and the optical conductivity $\sigma_{\gamma}(\omega)$ along the $\gamma(=x,y)$-direction of the two-leg ladders.
    Numerical results obtained by the Lanczos diagonalization method have clarified that the multi-triplon singlet states, including two- and three-triplon excitations, can be detected by observing these quantities.
    Furthermore, we have found that ladders with random potentials also have non-negligible contributions of these excitations to $\sigma_x(\omega)$.
}

\sloppy
\maketitle

%-----------------
\section{Introduction}
\label{sec1}

Quantum spin systems have long been regarded an ideal arena for exploring quantum many-body phenomena. In particular, $S=1/2$ two-leg spin ladders are notable for their tractability and typical low-dimensional features of quantum systems, such as disordered ground states, finite spin gaps, and non-trivial excitations~\cite{dagotto1,dagotto2}. The elementary excitations in $S=1/2$ two-leg ladders are quasiparticles called "triplons," each carrying a spin of $S=1$~\cite{triplon}.
The triplon state is regarded as a quasiparticle in a wide class of quantum spin systems with finite spin gaps, and its dynamical properties have been examined in various materials~\cite{NENP,CuGeO3,CuNO32,SrCu2BO32,TlCuCl3}. Also, in two-leg ladders, experimental studies using inelastic neutron scattering (INS) have successfully identified the finite energy gap and the single-triplon dispersion~\cite{INS_Sr14Cu24O41,La6Ca8Cu24O41}.

Recent experimental achievements strongly indicate that two-leg spin ladders are excellent platforms for studying two-triplon states; high-resolution analysis of INS experiments has confirmed the existence of a two-triplon continuum and bound states in the spin-triplet ($S=1$) sector of the excited states~\cite{La4Sr10Cu24O41,BPCB}. Besides these two-triplon states with $S=1$, which are detectable with INS, two-triplon singlet ($S=0$) excitations have also been observed using Raman spectroscopy~\cite{Sr15Cu24O41_raman,ladders_raman} and phonon-assisted optical absorption~\cite{phonon_assist_opt}. In addition, resonant inelastic x-ray scattering (RIXS) has been employed to explore two-triplon $S=0$ excitations~\cite{rixs_Sr14Cu24O41,rixs_Sr14Cu24O41_new}. These experimental results have been compared with theoretical studies on two-triplon states~\cite{zheng,schmidt,nunner,nocera,nagao,kumar}.

Turning our attention to recent studies of other quasiparticles in quantum magnets, we can see various experimental challenges to observe excited states having three or more numbers of quasiparticles, contributing to the spectral properties of these systems in high-energy regions.
The majority of these studies are devoted to bound states formed by strong inter-quasiparticle interactions and are investigated using INS~\cite{3magnon_NaMnO2,4magnon_FeI2}, THz spectroscopy~\cite{3spinon_string,6magnon_FeI2}, and infrared magnetospectroscopy~\cite{4magnon_FePSe3}.
Additionally, multi-quasiparticle continuum states composed of weakly interacting (or free) quasiparticles have also been examined in relation to the high-energy spectral components of INS or RIXS in 1D chains~\cite{4spinon_LiCuVO4,4spinon_Sr2CuO3}, or those of RIXS in a kagome antiferromagnet~\cite{4spinon_Kagome}.
As for the triplon states, in contrast, it has been only proposed theoretically that three-triplon states of the ladder can appear in the RIXS intensity~\cite{rixs_3tri}. Thus, methods to detect multi-triplon states are still lacking.

We present an alternative framework to investigate multi-triplon excitations of the two-leg ladders in the $S=0$ sector. Our approach is based on an analytical treatment for the one-dimensional (1D) ionic Hubbard model, a single-band Hubbard chain with an alternating potential~\cite{komaki}, where the charge disproportionation at each site is represented in terms of the spin operators of the corresponding Heisenberg model.
In this work, the analytical method is extended to Hubbard models on general lattices and applied to the two-leg Hubbard ladders with site-dependent potentials.
We have analytically proved that the spin-degrees of freedom of the two-leg ladders can have finite contributions to the dynamical charge structure factor $N(\bm{k},\omega)$ and the optical conductivity $\sigma_{\gamma}(\omega)$ along the $\gamma(=x,y)$-direction of the ladders.

Then, we numerically demonstrate that these spectral functions probe the two and three-triplon excitations of two-leg ladders with periodic potentials and with strong rung coupling, where spin dimers called ``rungs'' weakly couple along one direction.
The observed excitations can be selected by tuning the periodicity of the site-dependent potential or the wave number perpendicular to the leg.
Obtained results of $N(\bm{k},\omega)$ and $\sigma_{\gamma}(\omega)$ calculated with the Lanczos diagonalization elucidate that 
their spectral peaks lie in the two- and three-triplon continuum and the two-triplon singlet bound state.
Also, analytical results with bond operator method~\cite{sachdev,gopalan,normand} are found to be consistent with our calculated results.
Furthermore, we investigate the optical conductivity $\sigma_{x}(\omega)$ with random potential and show that observed spectral signatures result from these multi-triplon excitations.

This paper is organized as follows. 
In Sec.~\ref{sec_theory}, our theoretical framework for the single chain is extended to general lattices
and then applied to two-leg ladders.
In Sec.~\ref{sec_results}, our numerical and analytical results are presented, and Section~\ref{sec_conclusion} is devoted to the conclusion. 

%------------------------------------------------------------
\section{Theory}
\label{sec_theory}

\subsection{General Hubbard models}

Here, we explain the method, which is an extension of our previous framework for the 1D ionic Hubbard model~\cite{komaki}.
We employ the single-band Hubbard model at half-filling with a site-dependent potential.
The Hamiltonian is given by
\begin{align}
  {\cal H} =- \sum_{j l \tau} t_{jl} c^\dagger_{j\tau}c_{l\tau}+ U\sum_{l} n_{l\uparrow}n_{l\downarrow} 
  + \sum_{l} e_l (n_i-1),
  \label{eq_ionic}
\end{align}
where $c^{\dagger}_{j\tau}$ ($c_{j\tau}$) is the creation (annihilation) operator of an electron with spin $\tau$ at site $j$, $n_{j\tau}=c^{\dagger}_{j\tau}c_{j\tau}$, and $n_{j}=n_{j\uparrow}+n_{j\downarrow}$.
The parameter $U$ gives the on-site Coulomb interaction, and $t_{jl}$ represents the transfer integral between sites $j$ and $l$.
In addition, $e_j$ denotes the site-dependent potential at site $j$.
Here we set $t_{jj}=0$ and $e_j= \epsilon_j + \phi_j$ at site $j$,
where $\epsilon_j$ is an intrinsic site-dependent potential and $\phi_j$ is a virtually introduced scalar field.
For the large-$U$ case of this model, we derive the Heisenberg Hamiltonian as~\cite{katsura,komaki}
\begin{equation}
  {\cal H}^s = \sum_{jl} K_{jl} \left( \bm{S}_j \cdot \vb*{S}_{l} - \frac{1}{4} \right), \label{eq_Heisen}
\end{equation}
with the exchange interaction $K_{jl} = 2 t_{jl}^2/(U - e_j + e_{l})$.
The charge disproportionation $ \delta n_{l} \equiv n_{l} - 1 $ from the half-filling for $\phi_l = 0$ is evaluated as
\begin{align}
  \delta n^s_{l} = \left. \frac{\partial {\cal H}^s }{\partial \phi_l} \right|_{\phi_l=0} 
  =\sum_{j}  \eta_{lj}  \left( \bm{S}_l \cdot \bm{S}_{j} - \frac{1}{4} \right),
  \label{eq_delta_nl}
\end{align}
where $\bm{S}_j$ is the spin operator at site $j$, and the coefficient $\eta_{lj}$ associating $\delta n^s_{l}$ with the spin operators is given by
\begin{align}
  \eta_{lj} =  \frac{8 U t^2_{lj}  (\epsilon_l - \epsilon_j) }{ [ U^2 - (\epsilon_l - \epsilon_j)^2 ]^2}.
  \label{eq_eta}  
\end{align}
Here, the superscript $s$ of quantities means the contribution from the spin degrees of freedom.
%, and these quantities with the superscript $s$  are evaluated using the Heisenberg model derived here.

Next, let us consider the dynamical charge structure factor $N(\bm{k},\omega)$ of the Hubbard model~(\ref{eq_ionic}) defined by
\begin{equation}
  N(\bm{k},\omega) =  \frac{1}{N} \sum_{\alpha \ne 0} |\langle\alpha | n_{\bm k} |0\rangle|^2
  \delta(\omega - E_\alpha + E_0 ), \label{eq_nkw}
\end{equation}
where $n_{\bm k}=\sum_l e^{ -i {\bm k} \cdot \bm{R}_l} n_l$, ${\bm k}$ denotes the wave number, $N$ is the system size, and $\bm{R}_l$ represents the location of site $l$.
Its counterpart $N^s(\bm{k},\omega)$ of the spin degrees of freedom is obtained by replacing $n_{\bm k}$ in Eq.~(\ref{eq_nkw}) with
\begin{equation}
  \delta n^s_{\bm k}=\sum_l e^{ -i \bm{k}\cdot \bm{R}_l} \delta n^s_l
  = \sum_l e^{ -i \bm{k}\cdot \bm{R}_l} \sum_{j}  \eta_{lj}  \left( \bm{S}_l \cdot \bm{S}_{j} - \frac{1}{4} \right).
  \label{eq_dnk}
\end{equation}
At the same time, the eigenstate $\ket{\alpha}$ and the eigenenergy $E_\alpha$ of the Hamiltonian~(\ref{eq_ionic})
are also replaced by their counterparts, $|\alpha\rangle^s$ and $E_\alpha^s$, of the Heisenberg Hamiltonian~(\ref{eq_Heisen}).

Now, we discuss the optical conductivity of the Hubbard model~(\ref{eq_ionic}) given by
\begin{align}
  \sigma_\gamma(\omega ) 
  =\frac{\pi}{N\omega } \sum_{\alpha \ne 0} |\langle\alpha |J_\gamma |0\rangle|^2 
  \delta(\omega - E_\alpha + E_0)
\end{align}
and $\sigma^s_\gamma(\omega)$ in terms of the spin degrees of freedom,
where $J_\gamma$ is the $\gamma$ component of the current operator of the model~(\ref{eq_ionic}).
Following the procedure to derive $N^s(\bm{k},\omega)$, we represent $J_\gamma$ with respect to the operators of spin system~(\ref{eq_Heisen}).
To this aim, let us express the electric polarization~\cite{katsura} for $\gamma$ direction by using $\delta n^s_j$ of Eq.~(\ref{eq_delta_nl}) as
\begin{align}
  P^{s}_\gamma \equiv \sum_j (-e)R_j^\gamma \delta n^s_j
  = -e \sum_{ij} \eta_{ij} R^\gamma_i \left( \vb*{S}_i \cdot \vb*{S}_j -  \frac{1}{4} \right),
\end{align}
where $e$ is the absolute value of the charge of an electron.
Then the current operator is derived as~\cite{yokoi}
\begin{align}
  J^{s}_{\gamma}
  &\equiv  \frac{dP^s_{\gamma} }{dt} = -\frac{i}{\hbar} [ P^s_{\gamma}, {\cal H}^s] \nonumber \\
  &= -\frac{e}{\hbar} \sum_{ijl} \eta_{ij} \overline{K}_{il} ( R^\gamma_i - R^\gamma_j) \  \vb*{S}_i \cdot ( \vb*{S}_j \times \vb*{S}_l ),
  \label{eq_J_gamma}
\end{align}
where
\begin{align}
  \overline{K}_{il} =  \frac{4U(t_{il})^2 }{U^2 - (\epsilon_i - \epsilon_l )^2},
  \label{eq_kbar}
\end{align}
and the optical conductivity is defined as
\begin{align}
  \sigma^s_\gamma(\omega ) 
  =\frac{\pi}{N\omega } \sum_{\alpha \ne 0} | \, ^{s\,}\!\langle\alpha | J^s_\gamma |0\rangle^s |^2
  \delta(\omega - E^s_\alpha + E^s_0),
\end{align}
As $n^s_{\bm k}$ and $J^s_\gamma$ commute with the total spin $\vb*{S}_{\rm tot} =\sum_l \vb*{S}_l$,
the spectral functions $ N^s(\bm{k},\omega)$ and $\sigma^s_\gamma(\omega )$ can detect the singlet excitation~\cite{katsura,yokoi}.
Also, it is noted that these formulas can be applied to the 1D ionic Hubbard model by setting
\begin{align}
  t_{j j+1} = t, \ \epsilon_j = - (-1)^j \frac{\Delta}{2}, \ {\rm and} \ R_j = a j,
\end{align}
with the lattice constant $a$ and the strength of the potential $\Delta$. Then, it can be confirmed that the same results of Refs.~\citen{katsura,yokoi,komaki} are obtained.

We here stress the importance of the quantity $\eta_{lj}$ in our framework.
The point is that $\eta_{lj}$ appears both in $N^s(\bm{k},\omega)$ and $\sigma^s_\gamma(\omega )$ and that $\eta_{lj}$ is proportional to the energy difference $(\epsilon_l - \epsilon_j)$ between the neighboring sites $l$ and $j$. Therefore, the finite site-dependence of the potential $\epsilon_l$ is essential for observing these spectral functions.

%----------------------------------

\subsection{Two-leg ladders}

\begin{figure}
  \begin{center}
    \includegraphics[width=5.0cm]{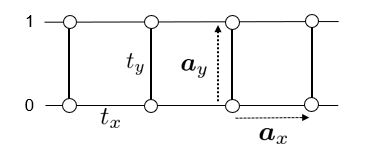}
    \caption{
      A Hubbard ladder with two legs 0 and 1, and the transfer integrals $t_x$ and $t_y$.
      The arrows are the primitive translational vectors $\vb*{a}_x$ and $\vb*{a}_y$ of this system.
    }
    \label{fig_ladder1}
  \end{center}
\end{figure}

Here, let us focus on two-leg ladders with only nearest-neighbor hoppings and periodic potentials.
Some notations are introduced to discuss the ladder systems in detail.
We label a site in a ladder by its location $\vb*{R}=(x,y)$ and use the primitive translational vectors $\vb*{a}_x$ and $\vb*{a}_y$.
Here, $\vb*{a}_{x(y)}$ is parallel to the legs (rungs), as shown in Fig.~\ref{fig_ladder1}.
A leg in a ladder is labeled by an integer $n=0,1$, and the location of a site in the leg $n$ is given by $\vb*{R}=(x,na_y)$,
where $a_\gamma=|\vb*{a}_\gamma|$ is the lattice constant along the $\gamma$-direction.
In addition, a site $\vb*{R}$ in the leg $n$ is represented by the short form $\vb*{R} \in \text{leg} \ n$.
Then the transfer integral $t_{\vb*{R}\vb*{R}'}$ between $\vb*{R}$ and $\vb*{R}'$ is given as
\begin{align}
  t_{\vb*{R}\vb*{R}'} =
  \begin{cases}
    t_x  & \vb*{R}'=\vb*{R} \pm \vb*{a}_x \\
    t_y  &  \quad ( \vb*{R} \in  \text{leg} \ 0 \  \text{and}  \ \vb*{R}' = \vb*{R} + \vb*{a}_y ) \\
    &  \text{or}  \  (\vb*{R} \in \text{leg}\ 1 \ \text{and} \ \vb*{R}'= \vb*{R} - \vb*{a}_y) \\        
    0    &  \text{otherwise}
  \end{cases}
  \label{eq_hopp_2leg}
\end{align}
and the site-dependent potential is represented by
\begin{equation}
  \epsilon_{\bm R} = -\frac{\Delta}{2} e^{-i \bm{Q}\cdot\bm{R}},
  \label{eq_ptl_2leg}
\end{equation}
where $\Delta>0$ gives the strength of the potential and $\vb*{Q}=(Q_x, Q_y)$ is a wave vector characterizing the periodicity.
Here the $\gamma$-component $Q_\gamma$ is limited to $Q_\gamma=0$ or $\pi/a_\gamma$.
Then the energy difference $\epsilon_l - \epsilon_j$ in Eq.~(\ref{eq_eta}) for a neighboring pair $(l,j)$ with $(\vb*{R}_l,\vb*{R}_j)=(\vb*{R},\vb*{R}\pm\vb*{a}_\gamma)$ along the $\gamma$-direction is evaluated as
\begin{align}
  \epsilon_{\vb*{R}} - \epsilon_{\vb*{R}\pm\vb*{a}_\gamma}
  &= - e^{-i \bm{Q}\cdot\bm{R}}\frac{\Delta ( 1 - e^{ \mp i \bm{Q}\cdot\bm{a}_\gamma} )}{2} \nonumber \\
  &=   - e^{-i \bm{Q}\cdot\bm{R}} \Delta_{\gamma}(\vb*{Q}),
\end{align}
where
\begin{align}
  \Delta_{\gamma}(\vb*{Q}) \equiv
  \begin{cases}
    \Delta  & Q_\gamma= \pi/a_\gamma \\
    0  & Q_\gamma=0
  \end{cases}
\end{align}
and the corresponding $\eta_{lj}$ is simplified as
\begin{align}
  \eta_{\vb*{R}\vb*{R}\pm\vb*{a}_\gamma} (\vb*{Q})=
  -e^{-i \bm{Q}\cdot\bm{R}} \eta_\gamma (\vb*{Q}),
  \label{eq_eta_2leg}
\end{align}
where
\begin{align}
  \eta_\gamma (\vb*{Q}) \equiv  \frac{8 U t^2_{\gamma}  \Delta_{\gamma} (\vb*{Q}) }{ [ U^2 - \Delta^2_{\gamma} (\vb*{Q}) ]^2}.
  \label{eq_def_eta}
\end{align}

In the same manner,
$\overline{K}_{il}$ of Eq.~(\ref{eq_kbar}) for a pair $(i,l)$ with $(\vb*{R}_i,\vb*{R}_l)=(\vb*{R},\vb*{R}\pm\vb*{a}_\gamma)$
is obtained as
\begin{align}
  \overline{K}_{\vb*{R}\vb*{R}\pm\vb*{a}_\gamma} (\vb*{Q})=
  \frac{4 U t^2_{\gamma} }{ U^2 - \Delta^2_{\gamma} (\vb*{Q})} \equiv \overline{K}_{\gamma} (\vb*{Q}).
\end{align}
In the following, the argument $\vb*{Q}$ of $\eta_\gamma (\vb*{Q})$ and $\overline{K}_{\gamma} (\vb*{Q})$ is omitted unless necessary.
Here, it is valuable to note that $\overline{K}_{\gamma}$ is equal to the exchange interaction of the two-leg spin ladder whose Hamiltonian of Eq.~(\ref{eq_Heisen}) is represented as
\begin{equation}
  {\cal H}^s = \overline{K}_{x}  \sum_{\vb*{R}} \bm{S}_{\vb*{R}} \cdot \bm{S}_{\vb*{R}+\vb*{a}_x } 
  +  \overline{K}_{y} \sum_{\vb*{R}  \in \text{leg}\ 0}\bm{S}_{\vb*{R}} \cdot \bm{S}_{\vb*{R}+\vb*{a}_y }.
  \label{eq_Heisen_ladder}
\end{equation}

Using these results, $\delta n^s_{\bm k}$ of a two-leg ladder is expressed as
\begin{align}
  \delta n^s_{\vb*{k}} = {\mathcal O}_x( {\bm k}' ) + {\mathcal O}_y({\bm k}' ), 
  \label{eq_dnk2}  
\end{align}
where
\begin{align}
  {\mathcal O}_x({\bm k}' ) &\equiv
  - \eta_{x} \sum_{\bm R} e^{-i \vb*{k}'\cdot\vb*{R} }
  \sum_{\vb{r} = \pm \vb*{a}_x } \vb*{S}_{\vb*{R}} \cdot \vb*{S}_{\vb*{R}+\vb*{r}},
  \label{eq_dNx}
  \\
    {\mathcal O}_y({\bm k}' ) &\equiv
    - \eta_{y} \sum_{\bm R} e^{-i \vb*{k}'\cdot\vb*{R} }
    \vb*{S}_{\vb*{R}} \cdot \vb*{S}_{\vb*{R}+\vb*{a}_y},
    \label{eq_dNy}  
\end{align}
and 
\begin{align}
  \vb*{k}' \equiv \vb*{k} +\vb*{Q}.
  \label{eq_kdash}    
\end{align}
Here, the constant (-1/4) is omitted because it has no contribution to $N^s(\bm{k},\omega)$.
The derivation of Eq.~(\ref{eq_dnk2}) is shown in Appendix A.
These results show that $N^s(\bm{k},\omega)$ probes spin excitations with the wave number $\bm{k}'$.
Therefore we represent and plot $N^s(\vb*{k},\omega)$ as a function of $\vb*{k}'$ in the following.
This is the generalization of the property of the 1D ionic Hubbard model, where $N^s(k,\omega)$ probes excited states with $k'=k+\pi$~\cite{komaki}.

Next, consider the current operator~(\ref{eq_J_gamma}).
Following the same procedure as that of $\delta n^s_{\vb*{k}}$, $J^s_x$ is simplified as 
\begin{align}
  &J^s_x
  = -\frac{\eta_x a_x e}{\hbar} 
  \sum_{\vb*{R}} e^{-i\vb*{Q}\vdot\vb*{R}}
  \left[
    2\overline{K}_{x} \vb*{S}_{\vb*{R}} \cdot ( \vb*{S}_{\vb*{R}+\vb*{a}_x} \times \vb*{S}_{\vb*{R}-\vb*{a}_x}) \right.  \nonumber \\
    &+\overline{K}_{y} \vb*{S}_{\vb*{R}} \cdot ( \vb*{S}_{\vb*{R}+\vb*{a}_x} \times \vb*{S}_{\vb*{R}+\vb*{a}_y} )
    \left.
    - \overline{K}_{y} \vb*{S}_{\vb*{R}} \cdot ( \vb*{S}_{\vb*{R}-\vb*{a}_x} \times \vb*{S}_{\vb*{R}+\vb*{a}_y} )
    \right].
  \label{eq_Jx_2leg}
\end{align}
This tells us that $\sigma^s_x(\omega )$ probes excited states with $\vb*{Q}$.
For the $y$-direction, we need to be more careful. The current operator is evaluated as
\begin{align}
  J^s_y= -\frac{\eta_y a_y e}{\hbar}   \overline{K}_{x}
  \sum_{\vb*{R}} \sum_{\vb*{r} = \pm \vb*{a}_x}
  e^{-i\widetilde{ \vb*{Q} }\vdot\vb*{R}}
  \vb*{S}_{\vb*{R}} \cdot ( \vb*{S}_{\vb*{R}+\vb*{a}_y} \times \vb*{S}_{\vb*{R}+\vb*{r}}),
  \label{eq_Jy_2leg}
\end{align}
where
\begin{align}
  \widetilde{\vb*{Q}}=\vb*{Q}+\left(0, \pi/a_y \right).
  \label{eq_tilde_K}
\end{align}
These relations suggest that $\sigma^s_y(\omega )$ probes excited states of spin systems with $\widetilde{\vb*{Q}}$, not $\vb*{Q}$.
The additional wave number $\pi/a_y$ results from the boundary conditions in the $y$-direction;
for $\vb*{R} \in$ leg 1, $\vb*{R} + \vb*{a}_y = \vb*{R} - \vb*{a}_y$.
It should also be noted that
\begin{align}
  {\mathcal O}_x({\bm k}' )  =  J^s_x= 0 & \ \text{for} \ Q_x =0 \ \text{and} \nonumber \\
  {\mathcal O}_y({\bm k}' )  =  J^s_y= 0 & \ \text{for} \ Q_y =0,
  \label{eq_zero_op}
\end{align}
which are deduced from Eq.~(\ref{eq_def_eta}) and the definitions of the operators.
The derivation of Eqs.~(\ref{eq_Jx_2leg}) and (\ref{eq_Jy_2leg}) is also explained in Appendix A.

%------------------------------------------------------------------------
\section{Results}
\label{sec_results}

\begin{figure}
  \begin{center}
    \includegraphics[width=4.5cm]{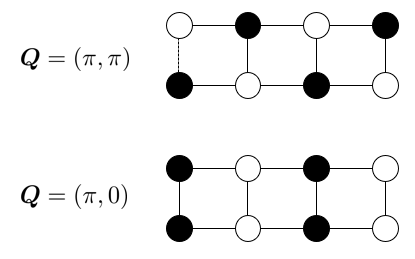}
    \caption{
      Two-leg ladders with the periodic potentials characterized by $\vb*{Q}=(\pi,\pi)$ and $\vb*{Q}=(\pi,0)$.
      The filled and open circles represent the periodic potentials.
    }
    \label{fig_ladder}  
  \end{center}
\end{figure}

Before discussing the details of our results, we would like to comment on our numerical calculation.
We set $e$, $\hbar$, $a_x$, and $a_y$ to unity for simplicity, and we limit $\bm{Q}$ in two cases; $\bm{Q}=(\pi,\pi)$ and $\bm{Q}=(\pi, 0)$  shown in Fig.~\ref{fig_ladder}.
For the spectral functions, the $\delta$-function is replaced with the Lorentzian with finite broadening $\epsilon=0.01t_x$, and we impose the periodic boundary condition along the $x$-direction unless otherwise noted.
For numerical calculations of physical quantities introduced here, we use Lanczos diagonalization.
We also note that quantities with superscript $s$ are those of the two-leg spin ladders. Quantities without superscripts are from the Hubbard ladders.

\begin{figure}
  \begin{center}
    \includegraphics[width=6.0cm]{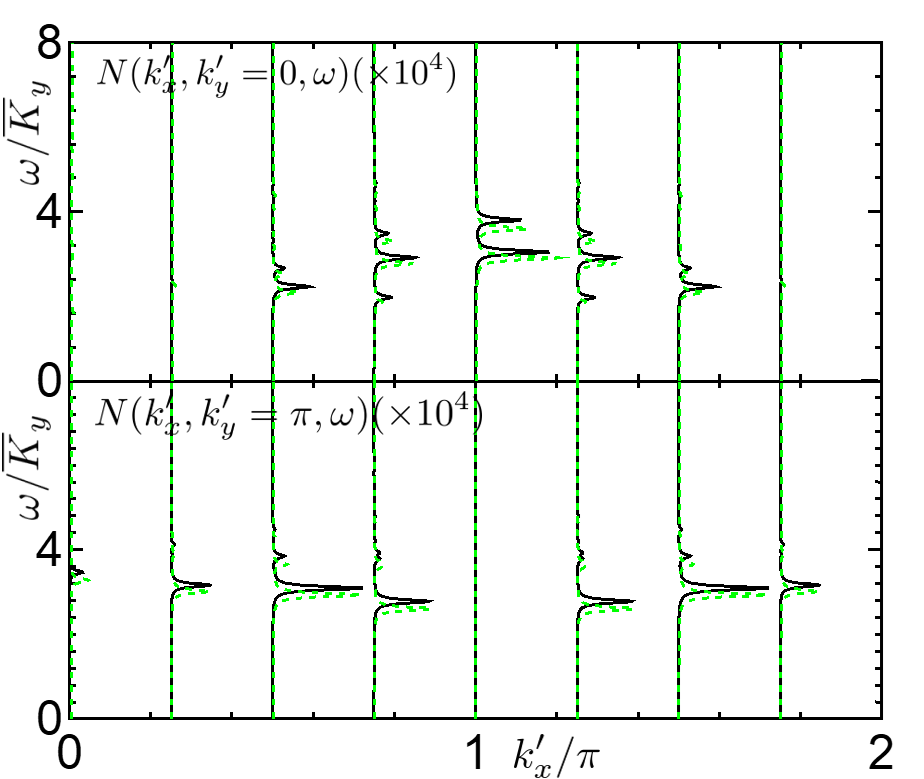}    
    \caption{
      (Color online) dynamical charge structure factors, $N(\vb*{k}',\omega)$ (green dashed lines) of the two-leg Hubbard ladder and $N^s(\vb*{k}',\omega)$ (black solid lines) of the spin ladder with
      $\vb*{Q}=(\pi,\pi)$, $t_y/t_x=1$, $U/t_x=20$, $\Delta/t_x=2$, and $N=16$.
    }
    \label{fig_nkw}  
  \end{center}
\end{figure}

\begin{figure}
  \begin{center}
    \includegraphics[width=6.0cm]{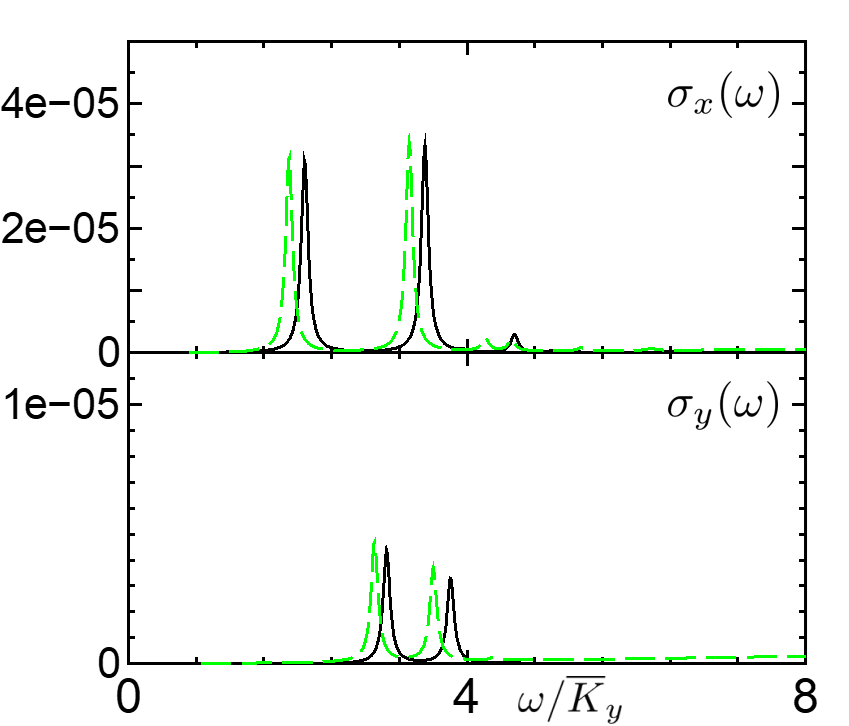}    
    \caption{
      (Color online) Optical conductivity $\sigma_\gamma(\omega)$ (green dashed lines) of the two-leg Hubbard ladder and $\sigma^s_\gamma(\omega)$ of the spin ladder for $\gamma=x,y$
      (black solid lines) with $\vb*{Q}=(\pi,\pi)$, $t_y/t_x=1$, $U/t_x=20$, $\Delta/t_x=2$, and $N=16$.
      The background resulting from the charge degrees of freedom is subtracted from the result of $\sigma_\gamma(\omega)$.
    }
    \label{fig_sw}  
  \end{center}
\end{figure}

First of all, we examine that $N^s(\vb*{k}',\omega)$ and $\sigma^s_\gamma(\omega)$ of the spin ladder well reproduce their counterparts $N(\vb*{k}',\omega)$ and $\sigma_\gamma(\omega)$ of the Hubbard ladder in the low-energy region.
Figure~\ref{fig_nkw} shows calculated results of $N(\vb*{k}',\omega)$ and $N^s(\vb*{k}',\omega)$ for the ladder with $\vb*{Q}=(\pi,\pi)$, $t_y/t_x=1$, $U/t_x=20$, $\Delta/t_x=2$, and $N=16$.
It can be seen that $N^s(\vb*{k}',\omega)$ well reproduce the peak structure of $N(\vb*{k}',\omega)$ for each $\vb*{k}'$.
The optical conductivity spectra $\sigma^s_\gamma(\omega)$ for $\gamma=x,y$ and their counterpart $\sigma_\gamma(\omega)$,
plotted in Fig.~\ref{fig_sw}, also demonstrate that the contributions from the spin degrees of freedom to these spectral functions well explain the low-energy behaviors of the original ones.

\begin{figure}
  \begin{center}
    \includegraphics[width=6.0cm]{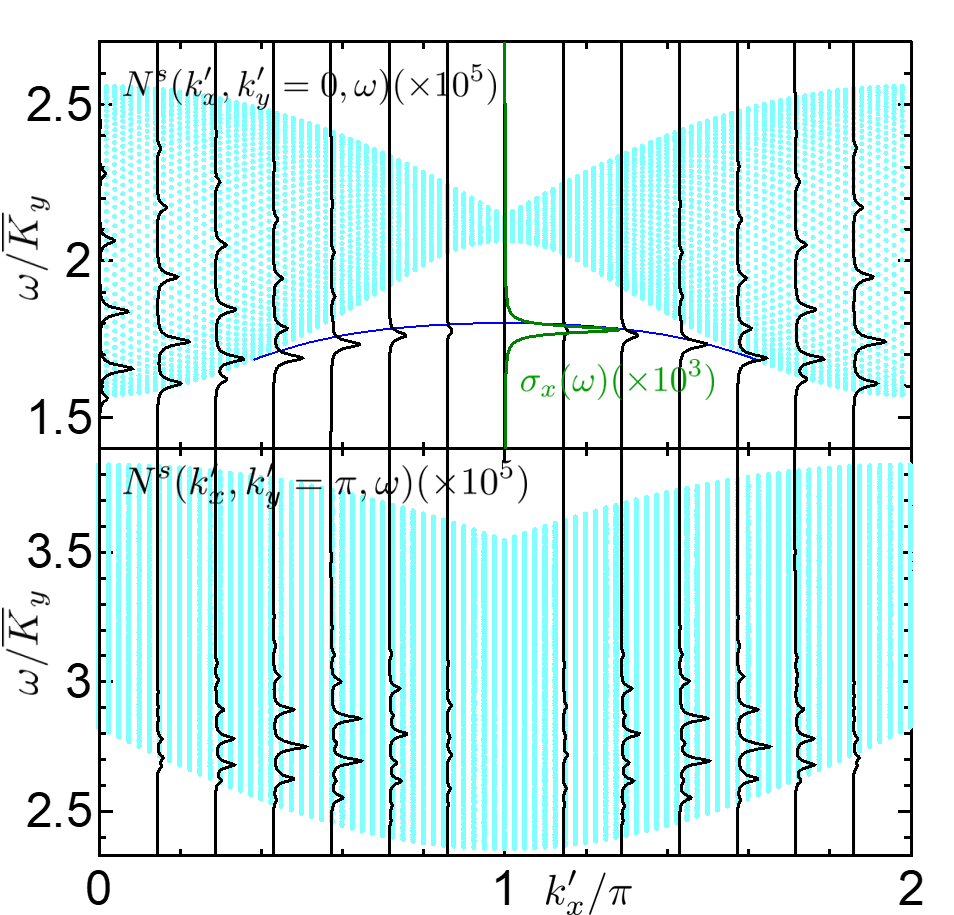}
    \caption{
      (Color online) dynamical charge structure factors $N^s(\vb*{k}',\omega)$ (black solid lines) and 
      optical conductivity $\sigma^s_x(\omega)$ (green solid line) of the two-leg spin ladder
      with $\vb*{Q}=(\pi,0)$, $t_y/t_x=2$, $U/t_x=20$, $\Delta/t_x=2$, and $N=28$.
      Light blue regions show the two-triplon continuum for $k'_y=0$ and the three-triplon continuum for $k'_y=\pi$.
      The blue solid line shows the dispersion of the two-triplon singlet bound state.
    }
    \label{fig_nkw3}  
  \end{center}
\end{figure}

Let us discuss how the multi-triplon excited states appear in these spectral functions.
For this purpose, we consider ladders with large $t_y$, where the series expansion method works well to evaluate the dispersion curves of multi-triplon excited states~\cite{zheng,oitmaa}.
Figure~\ref{fig_nkw3} displays $N^s(\vb*{k}',\omega)$ of the ladder with 
$\vb*{Q}=(\pi,0)$, $t_y/t_x=2$, $U/t_x=20$, $\Delta/t_x=2$, and $N=28$.
In addition, spectra of the multi-triplon states are also plotted.
For details on these multi-triplon states, please see Appendix B.
In the upper panel of this figure, we compare $N^s(k'_x, k'_y=0,\omega)$ with the dispersions of the two-triplon continuum and the two-triplon singlet bound state.
The result of $\sigma_x^s(\omega)$ is also shown at the wave number $(\pi, 0)$ because $\sigma_x(\omega )$ probes excited states with $\vb*{Q}$ as discussed before.
These results demonstrate that $N^s(k'_x, k'_y=0,\omega)$ and $\sigma_x^s(\omega)$ well observe the above-mentioned two-triplon excitations, including the continuum and the singlet bound state. 
In addition, a significant part of the spectral weight of $N^s(k'_x, k'_y=0, \omega)$ lies close to the lower boundary of the 2-triplon continuum.
In contrast, $N^s(k'_x, k'_y=\pi,\omega)$, plotted in the lower panel of Fig.~\ref{fig_nkw3}, is not so concentrated near the lower boundary of the three-triplon continuum; its spectral weight lies mainly in two elliptical regions, one centered at $(k'_x/\pi,\omega/\overline{K}_y)=(0.5,2.8)$ and the other at  $(k'_x/\pi,\omega/\overline{K}_y)=(1.5,2.8)$.
Here, we make a comment on the relation between our results and those of Ref.~\citen{rixs_3tri}; the cited paper theoretically shows that the three-triplon bound states formed by the three-body interaction have significant intensities in the RIXS spectrum of the two-leg ladder with $\overline{K}_x >\overline{K}_y$~\cite{rixs_3tri}.
On the other hand, $N^s(k'_x, k'_y=\pi,\omega)$ of our study does not detect the three-triplon bound state.
This is because the ladder treated here has a weaker $\overline{K}_x$ than $\overline{K}_y$, where the three-triplon bound states are not clearly visible in the spectral functions~\cite{rixs_3tri}.
Therefore, $N^s(k'_x,k'_y=\pi,\omega)$ would show some signature of the bound state for $\overline{K}_x > \overline{K}_y$, which is beyond the scope of this paper.

These behaviors of $N^s(\vb*{k}, \omega)$ are qualitatively reproduced using the bond operator method~\cite{sachdev,gopalan,normand}, where the spin operators on a rung are represented by boson operators, rung-singlet and rung-triplet bosons.
We use the procedure of Ref.~\citen{gopalan} for the two-leg ladder in order to evaluate $N^s(\vb*{k}',\omega)$ of the ladder with $\vb*{Q}=(\pi,0)$, and the results are shown in Fig.~\ref{fig_nkw_bondop}.
The analytical form of $N^s(\vb*{k}',\omega)$ is shown in Appendix C.
We can see that the spectral weight of $N^s(k'_x,k'_y=0,\omega)$ lies in the region enclosed between the lower and the upper boundaries of the two-triplon continuum.
In particular, the dominant part exists in the vicinity of the lower boundary.
It can also be seen that there is no signal of the bound state below the continuum, and a non-negligible amount of the weight is found near the upper boundary of the continuum.
This is because the bond operator theory of Ref.~\citen{gopalan} does not contain the inter-triplon interaction~\cite{komaki}.
The result of $N^s(k'_x,k'_y=\pi,\omega)$ also proves that the two ellipsoidal-shaped structures emerge in the three-triplon continuum.

We also find that $N^s(k'_x, k'_y=0,\omega)$ vanishes completely at $k'_x = \pi$, which is explained as follows.
Equation~(\ref{eq_zero_op}) shows that ${\mathcal O}_y( {\bm k}' )=0$ at $\vb*{Q}=(\pi, 0)$.
Thus $\delta n^s_{\vb*{k'}} = {\mathcal O}_x( {\bm k}' )$ in this case.
In addition, ${\mathcal O}_x( {\bm k}' )$ is reformulated as
\begin{align}
  {\mathcal O}_x({\bm k}' ) =   - \eta_{x} ( 1 + e^{-i k_x' } ) \sum_{\bm R} e^{-i \vb*{k}'\cdot\vb*{R} }
  \vb*{S}_{\vb*{R}} \cdot \vb*{S}_{\vb*{R}+\vb*{a}_x},
  \label{eq_dNx2}
\end{align}
which indicates that ${\mathcal O}_x({\bm k}' ) =0$ at $k_x = \pi$.
Hence $N^s(\vb*{k}',\omega)$ for $\vb*{Q}=(\pi,0)$ must be zero at $\vb*{k}'=(\pi,k_y)$ and around its vicinity $N^s(\vb*{k}',\omega)$ becomes small.
On the other hand, $\sigma_x^s(\omega)$ has a finite spectral weight at $\vb*{k}'=(\pi,0)$ and detects the two-triplon singlet bound states at this point.
The results of $N^s(\vb*{k}',\omega)$ and $\sigma_x^s(\omega)$ show that they complement each other well for observing the two-triplon singlet states.

\begin{figure}
  \begin{center}
    \includegraphics[width=7.0cm]{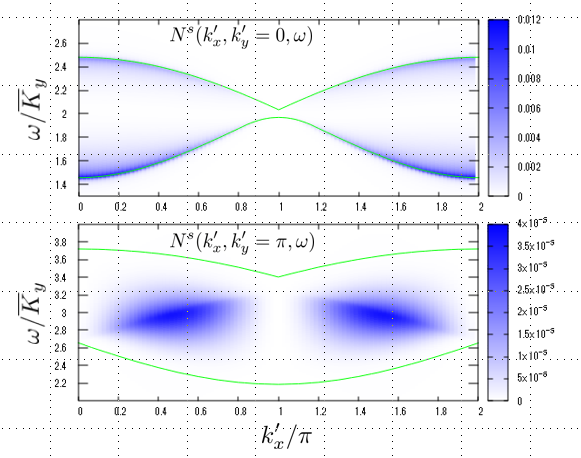}
    \caption{
      (Color online) $N^s(\vb*{k}',\omega)$ of the ladder with $\vb*{Q}=(\pi,0)$, $t_y/t_x=2$, $U/t_x=20$, and $\Delta/t_x=2$, calculated with the bond operator method.
      Green solid lines show upper and lower bounds of the two-triplon continuum for $k'_y=0$ and the three-triplon continuum for $k'_y=\pi$.
    }
    \label{fig_nkw_bondop}  
  \end{center}
\end{figure}

%------------------------------------------------------

\begin{figure}
  \begin{center}
    \includegraphics[width=6.0cm]{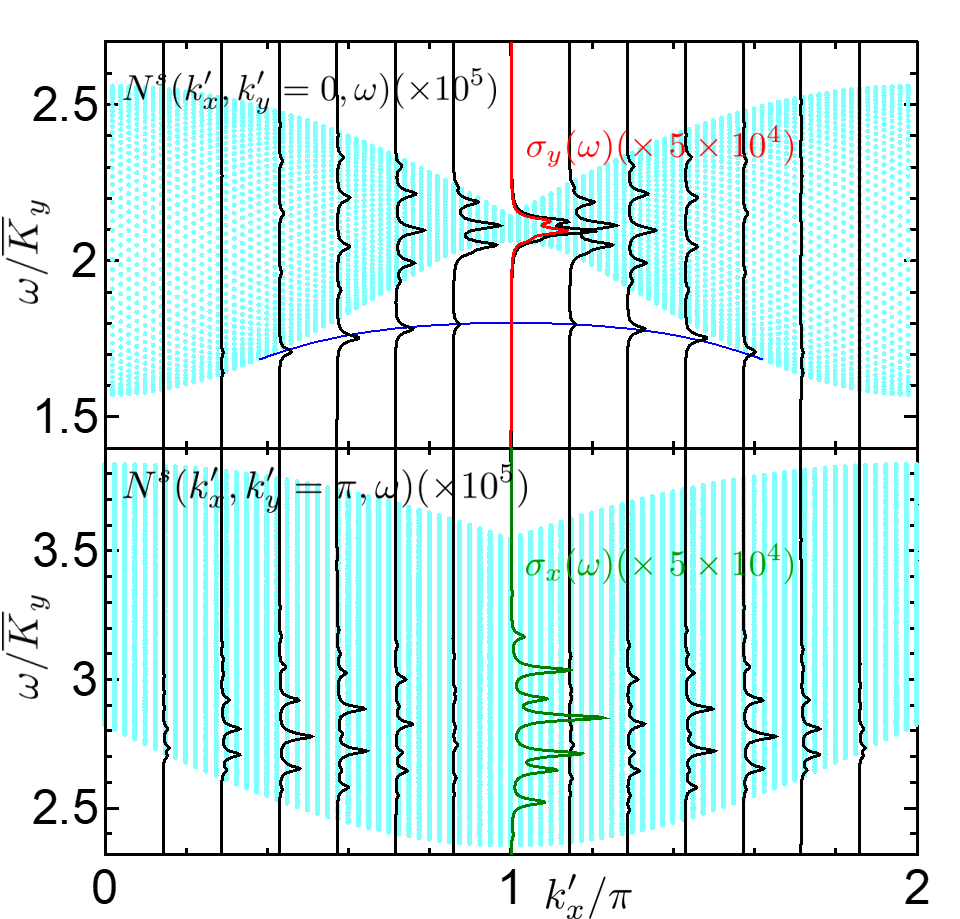}
    \caption{
      (Color online) $N^s(\vb*{k}',\omega)$ (black solid lines), $\sigma^s_x(\omega)$ (green solid line), and 
      $\sigma^s_y(\omega)$ (red solid line) of the two-leg spin ladder
      with $\vb*{Q}=(\pi,\pi)$, $t_y/t_x=2$, $U/t_x=20$, $\Delta/t_x=2$, and $N=28$.
      Light blue regions show the two-triplon continuum for $k'_y=0$ and the three-triplon continuum for $k'_y=\pi$.
      The blue solid line shows the dispersion of the two-triplon singlet bound state.
    }
    \label{fig_nkw2}
  \end{center}
\end{figure}

Next, we focus on the results of the ladder with $\vb*{Q}=(\pi,\pi)$.
Figure~\ref{fig_nkw2} shows $N^s(\vb*{k}',\omega)$, $\sigma^s_x(\omega)$,  $\sigma^s_y(\omega)$, and dispersions of two-triplon and three-triplon continuum for $\vb*{Q}=(\pi,\pi)$.
As shown in Eqs.~(\ref{eq_Jy_2leg}) and (\ref{eq_tilde_K}), $\sigma_y(\omega )$ is plotted at $\widetilde{ \vb*{Q} } =(\pi,0)$, while $\sigma_x^s(\omega)$ is shown at $\vb*{Q} = (\pi, \pi)$.
Physical parameters except $\vb*{Q} $ are the same as those of Fig.~\ref{fig_nkw3}.
The obtained results demonstrate that the two-triplon and three-triplon excitations are detected using these spectral quantities.
Especially, it is noteworthy that by changing the wavenumber $\vb*{Q}$, one can select a multi-triplon state observable in terms of optical conductivity $\sigma_x^s(\omega)$.

It should be also found that $N^s(k'_x, k'_y=\pi,\omega)$ for $\vb*{Q}=(\pi,\pi)$ is quite similar to that of $\vb*{Q}=(\pi,0)$,
although $N^s(k'_x, k'_y=0,\omega)$ of $\vb*{Q}=(\pi,\pi)$ is clearly different from that of $\vb*{Q}=(\pi,0)$.
The former similarity at $k'_y=\pi$ results from the fact that ${\mathcal O}_y({\bm k}')=0$ for $k'_y=\pi$ since
\begin{align}
  {\mathcal O}_y({\bm k}' ) = 
  - \eta_{y} (1 + e^{ i k'_y } )\sum_{ {\bm R} \in \  \text{leg} \ 0} e^{-i \vb*{k}'\cdot\vb*{R} }
  \vb*{S}_{\vb*{R}} \cdot \vb*{S}_{\vb*{R}+\vb*{a}_y}.
  \label{eq_dNy2}  
\end{align}
Hence, for $k'_y=\pi$ we obtain
\begin{align}
  \delta n^s_{\vb*{k}} = {\mathcal O}_x( {\bm k}' ) 
\end{align}
and thus $N^s(k'_x, k'_y=\pi,\omega)$ for $\vb*{Q}=(\pi,\pi)$ shows almost the same $(\vb*{k}',\omega)$-dependence as that of $N^s(k'_x, k'_y=\pi,\omega)$ for $\vb*{Q}=(\pi,0)$.

The latter difference of $N^s(k'_x, k'_y=0,\omega)$ is a little bit complicated;
the spectral weight of $N^s(k'_x, k'_y=0,\omega)$ for $\vb*{Q}=(\pi,\pi)$ is concentrated around $k'_x = \pi$ while $N^s(k'_x, k'_y=0,\omega)$ for $\vb*{Q}=(\pi,0)$ vanishes at $k'_x = \pi$.
This point originates from the finite ${\mathcal O}_y$ for $\vb*{Q}=(\pi,\pi)$.
In addition, $N^s(k'_x, k'_y=0,\omega)$ of $\vb*{Q}=(\pi,\pi)$ has a small contribution around $k'_x =0$, which is understood as follows.
It can be found that $\Delta_{\gamma} (\vb*{Q}) = \Delta$ for $\vb*{Q}=(\pi,\pi)$ and thus 
\begin{align}
  \eta_\gamma (\vb*{Q}) =   \frac{8 U t^2_{\gamma}  \Delta }{ [ U^2 - \Delta ]^2} = 2\Delta U \overline{K}_\gamma.
\end{align}
Consequently, we obtain 
\begin{align}
  \delta n^s_{\vb*{k}'=(0,0)} = - 4\Delta U  {\cal H}^s,
  \label{eq_dnk2_pipi}
\end{align}
which leads to $\langle \alpha| \delta n^s_{\vb*{k}'=(0,0)}|0 \rangle = 0$ for $|\alpha\rangle \ne |0\rangle$ and $N^s(k'_x=k'_y=0,\omega)=0$.

Now, we discuss how to observe these spectral quantities experimentally.
Unfortunately, not so many materials are thought to realize the Hubbard models with periodic potentials.
Besides TTF-CA and its derivatives~\cite{ttfca,ttfba,dmttf}, which are quasi-one-dimensional organic compounds with alternating potentials,
cold atom systems have been considered as promising realizations~\cite{messer}.
Although spectral quantities of cold atom systems are also observed experimentally~\cite{yang,anderson}, 
their spectral resolution seems insufficient to detect the spectral weights associated with the spin degrees of freedom discussed here.

Here, we explore materials with randomness as alternative systems with site-dependent potentials.
As a typical example, it has been pointed out that an organic ladder material ($\alpha$-DT-TTF)$_2$[Au(mnt)$_2$] has a weak disorder due to the donor molecule $\alpha$-DT-TTF~\cite{silva,silva2}.
This molecule has two types of structures, called cis- and trans-forms, which are randomly arranged in the ladder material~\cite{silva}.
The difference in the structures naturally leads to the difference in the energy level, which can lead to randomness in the site-dependent potential in the form that two types of sites with different potential energies are randomly arranged.
Furthermore, X-ray irradiation has been known to introduce randomness into organic correlated systems~\cite{sasaki},
which may allow us to treat various organic ladders with random potentials.

\begin{figure}
  \begin{center}
    \includegraphics[width=6.0cm]{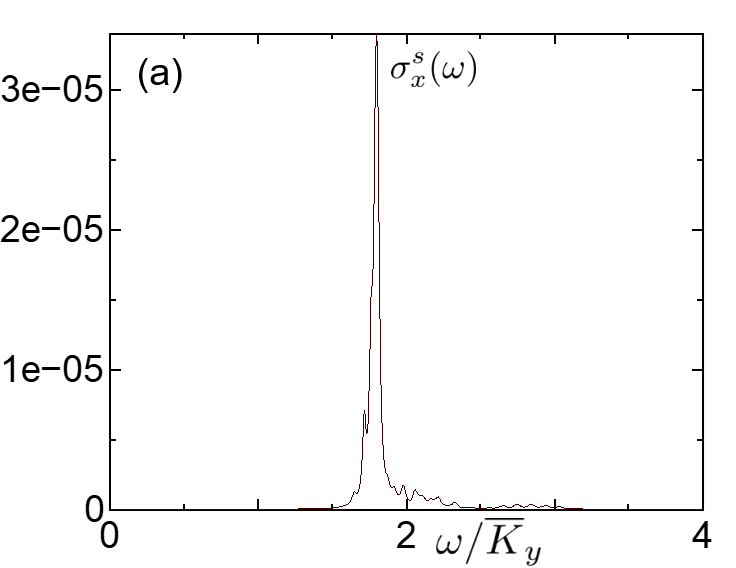}
    \includegraphics[width=6.0cm]{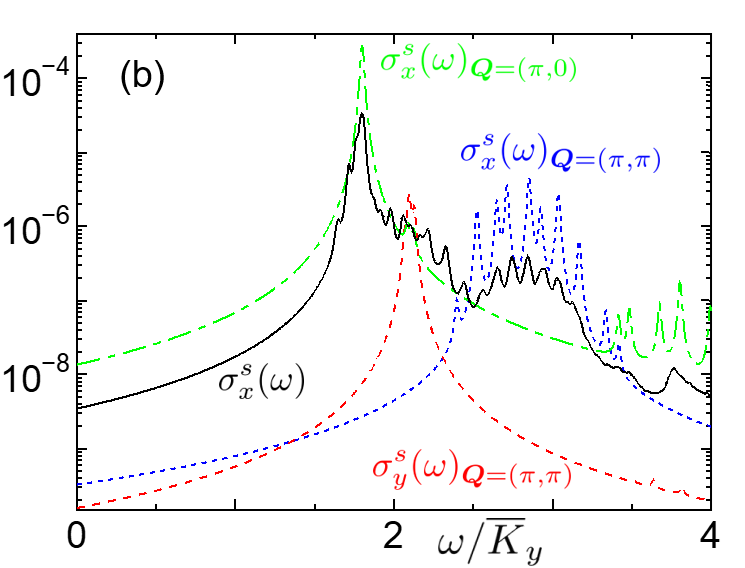}
    \caption{
      (Color online) (a) $\sigma^s_x(\omega)$ of the two-leg ladder with $t_y/t_x=2$, $U/t_x=20$, $\Delta/t_x=2$ and $N=28$ under the random potential.
      (b) The semi-log plot of $\sigma^s_x(\omega)$ shown in (a) (black solid lines).
      The results for ladders with periodic potentials, $\sigma^s_x(\omega)$ for $\vb*{Q}=(\pi,0)$ (green dot-dashed line),
      $\sigma^s_y(\omega)$ for $\vb*{Q}=(\pi,\pi)$ (red dashed line), and $\sigma^s_x(\omega)$ for $\vb*{Q}=(\pi,\pi)$ (blue dotted line) are also shown here.
      The wave number $\vb*{Q}$ for each quantity is depicted as the subscript.
    }
    \label{fig_sw_random}  
  \end{center}
\end{figure}

With ($\alpha$-DT-TTF)$_2$[Au(mnt)$_2$] in mind, we treat a two-leg ladder with the simple random potential expressed by
\begin{equation}
  \epsilon_l = -\frac{\Delta}{2} \xi_l,
  \label{eq_ptl_random}
\end{equation}
where $\xi_l$ is the random variable which takes 1 or -1 with the same probability.
Figure~\ref{fig_sw_random} (a) displays the calculated results of the optical conductivity $\sigma_x^s(\omega)$ of the ladder with the random potential.
The physical parameters are set to $t_y/t_x=2$, $U/t_x=20$, $\Delta/t_x=2$, $N=28$, and we take the average for 200 samples.
In addition, we use the same energy scale $\overline{K}_y =  8 U t^2_{y} / ( U^2 - \Delta )^2 $ as that of the case with the periodic potential of $\vb{Q}=(\pi,\pi)$ because of its usefulness although $\overline{K}_y$ is no longer the exact exchange interaction along the rung.
We have confirmed that the ladder with the random potential has the characteristic spectral structures in $\sigma_x^s(\omega)$;
there are a sharp peak at $\omega/\overline{K}_y \sim 1.8$, a shoulder structure at around $\omega/\overline{K}_y = 2.1$, 
and a small broad peak at around $\omega/\overline{K}_y=2.8$.
Judging from the energy scales where these structures appear, the sharp peak and shoulder are thought to originate from the two-triplon states, and the broad peak from the three-triplon states.
For further examination, the semi-log plot of the same data is shown in Fig.~\ref{fig_sw_random} (b), where numerical results of ladders with periodic potentials are also plotted.
Comparisons among these data suggest that the sharp peak stems from the two-triplon bound state and the shoulder from the continuum.
In addition, the broad peak corresponds to the three-triplon continuum.

\begin{figure}
  \begin{center}
    \includegraphics[width=6.0cm]{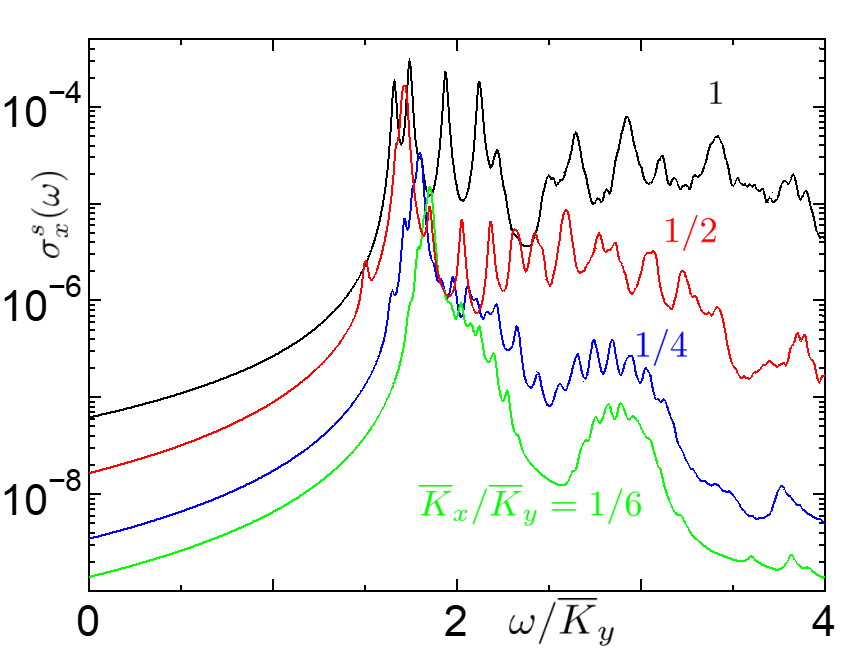}
    \caption{
      (Color online) optical conductivity $\sigma^s_x(\omega)$ of the two-leg ladders under the random potential with $U/t_y=10$, $\Delta/t_y=1$,  $N=28$, with increasing $\overline{K}_x/\overline{K}_y = (t_{x}/t_{y})^2$.
    }
    \label{fig_sw_random_Kx}  
  \end{center}
\end{figure}

Whether the spectral signature of the three-triplon state is well separated from that of the two-triplon states depends strongly on the competition between two energy scales, $\overline{K}_y$ and $\overline{K}_x$.
The former is roughly equal to the creation energy of one triplon, and the latter is proportional to the bandwidth of the triplon band.
Figure~\ref{fig_sw_random_Kx}, which shows $\sigma^s_x(\omega)$ for $U/t_y=10$, $N=28$, and $\Delta/t_y=1$, demonstrate how these spectral structures change with increasing $\overline{K}_x/\overline{K}_y = (t_{x}/t_{y})^2$.
For small $\overline{K}_x/\overline{K}_y$, the peak of the three-triplon band around $\omega \sim 3 \overline{K}_y$ is far from that of the two-triplon band for $\omega \sim 2 \overline{K}_y$.
As $\overline{K}_x/\overline{K}_y$ increases, the shoulder structure at $\omega/\overline{K}_y \sim 2$ originating from the two-triplon continuum becomes larger and broader.
At $\overline{K}_x/\overline{K}_y=1/2$, the shoulder merges with the peak of the three-triplon continuum, and the resulting broad shoulder ranging from $\omega/\overline{K}_y \sim 2 $ to $\omega/\overline{K}_y \sim 3.5$ is reminiscent of the multi-triplon state.
At $\overline{K}_x/\overline{K}_y=1$, it becomes difficult to find out the signature of the three-triplon state.
Since the ladder material ($\alpha$-DT-TTF)$_2$[Au(mnt)$_2$] is considered to be $(t_{x}/t_{y})^2=0.54$,~\cite{silva} the broad shoulder structure state might be detected experimentally.
Furthermore, materials with smaller $\overline{K}_x/\overline{K}_y$, such as (C$_5$H$_{12}$N)$_2$CuBr$_4$ ($\overline{K}_x/\overline{K}_y \sim 1/4$)~\cite{BPCB,BPCB2}, irradiated with X-rays could be much better ones to observe these multi-triplon states.

%----------------------------------
\section{Summary}
\label{sec_conclusion}

In this work, we have studied the low-energy spectral properties of the two-leg Hubbard ladders with site-dependent potentials.
Our approach combining the analytical treatment for the Hubbard models with the numerical diagonalization demonstrated that the spectral functions $N^s(\bm{k}',\omega)$ and $\sigma^s_{\gamma}(\omega)$, the contributions of the spin degrees of freedom to the dynamical charge structure factor and the optical conductivity along the $\gamma=x,y$ direction, detect the two- and three-triplon excitations of the two-leg spin ladders.
The key ingredient to detect these multi-triplon excitations is the finite site-dependent potential of the ladder.
In our framework, this point is expressed in terms of the quantity $\eta_{lj}$, which appears in the definitions of $N^s(\bm{k}',\omega)$ and $\sigma^s_{\gamma}(\omega)$.
Numerical results obtained using the Lanczos diagonalization are consistent with the results of bond operator theory.
We have also examined that, by tuning the periodicity of the potential, $\sigma^s_{\gamma}(\omega)$ can select detectable excited states;
for example, $\sigma^s_{x}(\omega)$ probes either the two-triplon bound state or the three-triplon continuum.
Furthermore, we confirmed that the two-leg ladder with the random potential also has finite spectral weights in $\sigma^s_{x}(\omega)$ stemming from the two- and three-triplon states, which suggest that ladder materials with intrinsic or acquired randomness and strong intra-rung couplings are good candidates for experimentally observing and identifying these multi-triplon states.

\begin{acknowledgment}
  %For environments for acknowledgement(s) are available: \verb|acknowledgment|, \verb|acknowledgments|, \verb|acknowledgement|, and \verb|acknowledgements|.

\section*{Acknowledgment}  
%  \verb|Acknowledgment|
  This work was supported by a financial support from the University of Tsukuba, Pre-Strategic Initiatives 'Development Center for High-Function and High-Performance Organic-Inorganic Spin Electronics'.

\end{acknowledgment}

\appendix
\section{Operators of Two-leg Ladders}

%\subsection{Operators of two-leg ladders}

Here, we derive some relations between operators used in this work.
Now let us start with the derivation of Eqs.~(\ref{eq_dnk2}-\ref{eq_kdash}) for the two-leg ladders with the nearest-neighbor hoppings and the periodic potentials.
Since the summation in Eq.~(\ref{eq_dnk}) with respect to $l$ and $j$ is restricted to the nearest-neighbor pairs, we obtain
\begin{align}
  \delta n^s_{\bm k}=
  & - \eta_{x}\sum_{\bm R} e^{-i \vb*{k}'\cdot\vb*{R} }
  \sum_{\vb{r} = \pm \vb*{a}_x }  \vb*{S}_{\vb*{R}} \cdot \vb*{S}_{\vb*{R}+\vb*{r}}  \nonumber \\
  &- \eta_{y}\sum_{{\bm R} \in \text{leg} \ 0 } e^{-i \vb*{k}'\cdot\vb*{R} }
  \vb*{S}_{\vb*{R}} \cdot \vb*{S}_{\vb*{R}+\vb*{a}_y}  \nonumber \\
  & - \eta_{y}\sum_{{\bm R} \in \text{leg} \ 1 } e^{-i \vb*{k}'\cdot\vb*{R} }
  \vb*{S}_{\vb*{R}} \cdot \vb*{S}_{\vb*{R}-\vb*{a}_y},
\end{align}
where the coefficient $\eta_{lj}$ for the pair $(lj)$ is replaced by $\eta_{\vb*{R}\vb*{R}\pm\vb*{a}_\gamma}$ of Eq.~(\ref{eq_eta_2leg}), and the constant (-1/4) in Eq.~(\ref{eq_delta_nl}) is omitted as noted above.
Then using the periodic boundary condition for the $y$-direction, $\vb*{R} + \vb*{a}_y = \vb*{R} - \vb*{a}_y$ for $\vb*{R} \in$ leg 1,
the second and third terms are summarized as
\begin{align}
  \delta n^s_{\vb*{k}}
  & =  - \eta_{x} \sum_{\bm R} e^{-i \vb*{k}'\cdot\vb*{R} }
  \sum_{\vb{r} = \pm \vb*{a}_x } \vb*{S}_{\vb*{R}} \cdot \vb*{S}_{\vb*{R}+\vb*{r}} \nonumber \\
  & - \eta_{y} \sum_{\bm R} e^{-i \vb*{k}'\cdot\vb*{R} } \vb*{S}_{\vb*{R}} \cdot \vb*{S}_{\vb*{R}+\vb*{a}_y}.
\end{align}

As for the current operator $J^s_x$, Eq.~(\ref{eq_J_gamma}) for $\gamma=x$ is evaluated as
\begin{align}
  J^{s}_{x}
  & = \frac{e}{\hbar} \eta_x \overline{K}_x\left[
    \sum_{{\bm R} } e^{-i \vb*{Q}\cdot\vb*{R} }
    (-a_x)  \vb*{S}_{\vb*{R}} \cdot ( \vb*{S}_{\vb*{R}+\vb*{a}_x} \times \vb*{S}_{\vb*{R}-\vb*{a}_x})
    \right. \nonumber \\
    &+  \left.\sum_{{\bm R} } e^{-i \vb*{Q}\cdot\vb*{R} }
    (a_x)  \vb*{S}_{\vb*{R}} \cdot ( \vb*{S}_{\vb*{R}-\vb*{a}_x} \times \vb*{S}_{\vb*{R}+\vb*{a}_x})  \right] \nonumber \\
  &+ \frac{e}{\hbar} \eta_x \overline{K}_y\left[
    \sum_{{\bm R} \in \text{leg} \ 0 } e^{-i \vb*{Q}\cdot\vb*{R} }
    (-a_x)  \vb*{S}_{\vb*{R}} \cdot ( \vb*{S}_{\vb*{R}+\vb*{a}_x} \times \vb*{S}_{\vb*{R}+\vb*{a}_y})
    \right. \nonumber \\
    &+\sum_{{\bm R} \in \text{leg} \ 0 } e^{-i \vb*{Q}\cdot\vb*{R} }
    (a_x)  \vb*{S}_{\vb*{R}} \cdot ( \vb*{S}_{\vb*{R}-\vb*{a}_x} \times \vb*{S}_{\vb*{R}+\vb*{a}_y}) \nonumber \\
    &+\sum_{{\bm R} \in \text{leg} \ 1 } e^{-i \vb*{Q}\cdot\vb*{R} }
    (-a_x)  \vb*{S}_{\vb*{R}} \cdot ( \vb*{S}_{\vb*{R}+\vb*{a}_x} \times \vb*{S}_{\vb*{R}-\vb*{a}_y}) \nonumber \\
    &+\left.\sum_{{\bm R} \in \text{leg} \ 1 } e^{-i \vb*{Q}\cdot\vb*{R} }
    (a_x)  \vb*{S}_{\vb*{R}} \cdot ( \vb*{S}_{\vb*{R}-\vb*{a}_x} \times \vb*{S}_{\vb*{R}-\vb*{a}_y})  \right].
\end{align}
Now, using the same boundary condition for $y$-direction, we obtain
\begin{align}
  J^{s}_{x}
  & = -\frac{e}{\hbar} \eta_x a_x\sum_{{\bm R} } e^{-i \vb*{Q}\cdot\vb*{R} }
  \left[
    2 \overline{K}_x \vb*{S}_{\vb*{R}} \cdot ( \vb*{S}_{\vb*{R}+\vb*{a}_x} \times \vb*{S}_{\vb*{R}-\vb*{a}_x})
    \right. \nonumber \\
    & + \overline{K}_y \vb*{S}_{\vb*{R}} \cdot ( \vb*{S}_{\vb*{R}+\vb*{a}_x} \times \vb*{S}_{\vb*{R}+\vb*{a}_y}) \nonumber \\
    & - \left.  \overline{K}_y \vb*{S}_{\vb*{R}} \cdot ( \vb*{S}_{\vb*{R}-\vb*{a}_x} \times \vb*{S}_{\vb*{R}+\vb*{a}_y}) \right]    .
\end{align}

For $J^s_y$, we need a slightly different treatment.
The definition~(\ref{eq_J_gamma}) for $\gamma=y$ is evaluated as
\begin{align}
  J^s_y=
  &\frac{e\eta_y\overline{K}_{x}}{\hbar} \left[
    \sum_{\vb*{R} \in \text{leg} \ 0} e^{-i\vb*{Q}\vdot\vb*{R}}
    \sum_{\vb*{r} = \pm \vb*{a}_x}
    ( -a_y ) \vb*{S}_{\vb*{R}} \cdot ( \vb*{S}_{\vb*{R}+\vb*{a}_y} \times \vb*{S}_{\vb*{R}+\vb*{r}}) \right.
    \nonumber \\
    &
    \left.
    + \sum_{\vb*{R} \in \text{leg} \ 1} e^{-i\vb*{Q}\vdot\vb*{R}}
    \sum_{\vb*{r} = \pm \vb*{a}_x}
    ( a_y ) \vb*{S}_{\vb*{R}} \cdot ( \vb*{S}_{\vb*{R}+\vb*{a}_x} \times \vb*{S}_{\vb*{R} + \vb*{r}})
    \right].
  \label{eq_appendix_Jy1}
\end{align}
Here we introduce the modified wave number $\widetilde{\vb*{Q}}=\vb*{Q}+\left(0, \pi/a_y \right)$ of Eq.~(\ref{eq_tilde_K}), and then obtain
\begin{align}
  e^{ -i \widetilde{\vb*{Q}} \cdot \vb*{R}} = 
  \begin{cases}
    e^{ -i \vb*{Q}\cdot\vb*{R}}  & \text{for} \ \vb*{R} \  \in \ \text{leg} \ 0 \\
    -e^{ -i \vb*{Q}\cdot\vb*{R}}  & \text{for} \ \vb*{R}  \ \in \ \text{leg} \ 1
  \end{cases}.
\end{align}
Therefore, Equation~(\ref{eq_appendix_Jy1}) is summarized as
\begin{align}
  J^s_y= -\frac{\eta_y a_y e}{\hbar}   \overline{K}_{x}
  \sum_{\vb*{R}} \sum_{\vb*{r} = \pm \vb*{a}_x}
  e^{-i\widetilde{ \vb*{Q} }\vdot\vb*{R}}
  \vb*{S}_{\vb*{R}} \cdot ( \vb*{S}_{\vb*{R}+\vb*{a}_y} \times \vb*{S}_{\vb*{R}+\vb*{r}}).
\end{align}

\section{Dispersion of Multi-Triplon States}

In this part, we make some comments on the dispersion relations of multi-triplon states.
As for the one-triplon dispersion, we use the result of Ref.~\citen{oitmaa}.
The dispersion $\omega_1(k_x)$ of the one-triplon with the wave number $k_x$ along $x$-direction is given by
\begin{align}
  %\omega_1(k_x, k_y=\pi) = \overline{K}_{y} \sum_{n = 0}^8\sum_{m = 0}^8 a_{n,m} \lambda^{n} \cos( m k_x),
  \omega_1(k_x) = \overline{K}_{y} \sum_{n = 0}^8\sum_{m = 0}^8 a_{n,m} \lambda^{n} \cos( m k_x),  
\end{align}
where $\lambda = \overline{K}_x/\overline{K}_y$ is the expansion parameter and the coefficient $a_{n,m}$ is shown in Table III of Ref.~\citen{oitmaa}.

An eigenvalue of an $n$-triplon unbound state with the wave number $p_{\rm tot}$ is given by the following simple summation
\begin{align}
  \omega^C_{nt} (p_{\rm tot})  = \sum_{i=1,\cdots,n} \omega_1(p_i)
\end{align}  
under the limitation of $\sum_{i=1,\cdots,n} p_i = p_{\rm tot}$.
By taking all allowed combinations of $(p_1,\cdots,p_n)$, we obtain the $n$-triplon continuum at $k_x=p_{\rm tot}$.
Here, we note that one triplon carries the wave number $k_y = \pi$ along $y$-direction.
Hence $n$-triplon states have the wave number $k_y = n\pi$ (mod $2\pi$).

As for the two-triplon singlet bound state, we follow the procedure of Ref.~\citen{kumar};
we employ the results of Ref.~\citen{zheng} for the dispersion relation $\omega^S_{2t}(k_x)$ of the two-triplon singlet bound state, which is given by
\begin{align}
  \omega^S_{2t}(k_x) =
  &\overline{K}_{y}  \left[ 2-\frac{3}{2}\lambda + \frac{19}{16}\lambda^2 - \frac{9}{32}\lambda^3 \right. \nonumber \\
    &- \left(\frac{1}{2}\lambda - \frac{1}{8}\lambda^2 + \frac{51}{128}\lambda^3 \right) \cos(k_x) \nonumber \\
    &- \left(\frac{5}{16}\lambda^2 + \frac{21}{32}\lambda^3 \right) \cos(2k_x) \nonumber \\    
    & \left. - \frac{37}{128}\lambda^3 \cos(3k_x) \right].
  \label{eq_2tri_singlet}
\end{align}

\section{Bond Operator Theory}
\label{subsec_bond}

We here review the bond operator theory~\cite{sachdev,gopalan,normand},
which was developed to investigate the ground state phase diagram of $S=1/2$ two-dimensional frustrated Heisenberg model~\cite{sachdev}.
The application to the two-leg ladder was made by Gopalan {\it et al.}~\cite{gopalan}, and more excellent treatment of the ladder is discussed later~\cite{normand}.
Here, we follow the procedure of Ref.~\citen{gopalan} and pick up only essential points to reproduce our calculations for the convenience of the reader.

Let us introduce the bond operators. The $\alpha$-component of the spin operator of the site $\vb*{R}=(l, n)$ on the leg $n=0$ or $1$ is represented as
\begin{align}
  S^{\alpha}_{0 l } & \equiv
  S^{\alpha}_{\vb*{R}=(l, 0)}
  = \frac{1}{2}(s^\dagger_l t_{l\alpha} + t^\dagger_{l\alpha} s_l
  - \varepsilon_{\alpha\beta\gamma} t^\dagger_{l\beta}t_{l\gamma} )
\end{align}
and
\begin{align}
  S^{\alpha}_{1 l} &\equiv
  S^{\alpha}_{\vb*{R}=(l, 1)}
  = \frac{1}{2}( -s^\dagger_l t_{l\alpha} - t^\dagger_{l\alpha} s_l
  - \varepsilon_{\alpha\beta\gamma} t^\dagger_{l\beta}t_{l\gamma} ),
\end{align}
where $s_l$ and $t_{l\alpha} (\alpha=x,y,z)$ are bond operators for the singlet and triplet states of a rung labeled by $l$.
Here, we use the summation convention for repeated indices and the totally antisymmetric tensor $\varepsilon_{\alpha\beta\gamma}$.
These operators have bosonic statistics to satisfy the algebra of the spin operator $S^\alpha_{\vb*{R}}$.
In addition, they follow the hard-core condition
\begin{align}
  s^\dagger_l s_l + t^\dagger_{l\alpha}  t_{l\alpha}  = 1
  \label{eq_hard_core}
\end{align}
for each rung.

Using the bond operators, we rewrite the Hamiltonian of the two-leg ladder~(\ref{eq_Heisen_ladder}) as
\begin{align}
  {\cal H}^s = H_0 + \lambda H_1,  
  \label{eq_Ham_boson}
\end{align}
where
\begin{align}
  H_0  &= \sum_l \overline{K}_y \left( -\frac{3}{4}s^\dagger_l s_l + \frac{1}{4} t^\dagger_{l\alpha} t_{l\alpha} \right) \nonumber \\
  &-\sum_l \mu_l( s^\dagger_l s_l  + t^\dagger_{l\alpha} t_{l\alpha} -1) 
\end{align}
and
\begin{align}
  H_1  = \frac{\overline{K}_y}{2} \sum_l \left( t^\dagger_{l\alpha} t_{l+1\alpha} s^\dagger_{l+1} s_l  + t^\dagger_{l\alpha} t^\dagger_{l+1\alpha} s_{l} s_{l+1} + \text{H.c.}  \right).
\end{align}
Here we note that the chemical potential $\mu_l$ is introduced to impose the hard-core condition~(\ref{eq_hard_core})
and the quartic term with respect to $t^{(\dagger)}_l$ ($H_2$ in Ref.~\citen{gopalan} ) is omitted here following the theoretical treatment~\cite{gopalan}.

Now, the replacement $s^{(\dagger)}_l \to \langle s^{(\dagger)}_l \rangle = \bar{s}$ using the mean field $\bar{s}$ is introduced to diagonalize the Hamiltonian~(\ref{eq_Ham_boson}).
In addition, the chemical potential $\mu_l$ is replaced by the site-independent one $\mu$ because of the translational invariance of the ladder system.
Applying the Fourier transformation of the operations $t_{l\alpha} = N_x^{-1} \sum_{k} t_{k\alpha} e^{ikl}$, 
we obtain the mean-field Hamiltonian
\begin{align}
  H_m(\mu,\bar{s})
  =& N_x\left( -\frac{3}{4} \overline{K}_y \bar{s}^2 - \mu \bar{s}^2 + \mu \right) \nonumber \\ % - \frac{N}{2} \left(\frac{J}{4} - \mu \right) 
  &+\sum_{k}[ \Lambda_k  t^\dagger_{k\alpha} t_{k\alpha} + \Delta_k (t^\dagger_{k\alpha}t^\dagger_{-k\alpha} +t_{k\alpha}t_{-k\alpha} )  ]
\end{align}
where
\begin{align}
  \Lambda_k = \frac{1}{4} - \mu + \lambda \overline{K}_y  \bar{s}^2 \cos k, \quad
  \Delta_k = \frac{1}{2 } \overline{K}_ y  \bar{s}^2 \cos k,
\end{align}
and $N_x$ is the number of rungs.
Using the Bogoliubov transformation
\begin{align}
  \gamma_{k\alpha} = \cosh \theta_k t_{k\alpha} + \sinh \theta_k t^\dagger_{-k\alpha}
\end{align}
with 
\begin{align}
  \cosh \theta_k &= \frac{1}{\sqrt{2} } \sqrt{ \frac{\Lambda_k}{\omega_k} +1 }
\end{align}
and
\begin{align}    
  \sinh \theta_k &= \frac{1}{\sqrt{2} } \text{sgn} ( \Delta_k ) \sqrt{ \frac{\Lambda_k}{\omega_k} - 1 },
\end{align}
we diagonalize the mean-field Hamiltonian as
\begin{align}
  H_m(\mu,\bar{s}) =& N_x\left( -\frac{3}{4} \overline{K}_y \bar{s}^2 - \mu \bar{s}^2 + \mu \right) - \frac{N_x}{2} \left(\frac{\overline{K}_y}{4} - \mu \right) \nonumber \\
  & +\sum_{k} \omega_k \left(  \gamma^\dagger_{k\alpha} \gamma_{k\alpha} + \frac{1}{2} \right).
\end{align}
Here, the quantity
\begin{align}
  \omega_k =\sqrt{\Lambda^2_k - (2\Delta_k)^2}
\end{align}
is the dispersion of the Bogoliubov boson, which corresponds to the triplon of the two-leg ladder.

The next step is to obtain physical quantities within this mean-field theory.
Following the method in the original paper~\cite{gopalan}, we calculate the mean fields $\bar{s}$ and $\mu$ at the ground state by numerically solving the saddle-point equations
\begin{align}
  \left\langle \frac{\partial H_m}{\partial \mu } \right\rangle = 0 \ \text{and} \  \left\langle \frac{\partial H_m}{\partial \bar{s} } \right\rangle = 0.
\end{align}
Then, we evaluate the dynamical charge structure factor $N^s(\bm{k}',\omega)$ for $\vb*{Q}=(\pi,0)$, where the charge disproportionation $\delta n^s_{{\bm k}'}$ of Eq.~(\ref{eq_dnk2}) is given by
\begin{align}
  \delta n^s_{\vb*{k}}
  &= {\mathcal O}_x( {\bm k}' ) \nonumber \\
  &=  - \eta_{x} \sum_{\bm R} e^{-i \vb*{k}'\cdot\vb*{R} }
  \sum_{\vb{r} = \pm \vb*{a}_x } \vb*{S}_{\vb*{R}} \cdot \vb*{S}_{\vb*{R}+\vb*{r}}.
\end{align}
For this purpose, we represent
\begin{align}
  \delta n^s_{{\vb*{k}}'}
  &= \delta n^0_{k_x'}     \quad \text{for}  \quad \bm{k}'=(k_x', 0) \\
  &= \delta n^{\pi}_{k_x'}  \quad \text{for}  \quad \bm{k}'=(k_x', \pi),
\end{align}
and then express these quantities with the boson operators.

First, we deal with
\begin{align}
  \delta n^0_{k_x'}
  =  - \eta_{x} \sum_{\bm l} e^{-i k_x' l } \sum_{r = \pm 1}( \vb*{S}_{0l} \cdot \vb*{S}_{0 l+r} + \vb*{S}_{1l} \cdot \vb*{S}_{1 l+r}  ),
\end{align}
which is necessary to calculate $N^s(k'_x, k'_y=0,\omega)$.
Here, we use the boson representation
\begin{align}
  &\bm{S}_{0l} \cdot \bm{S}_{0j}  +   \bm{S}_{1l} \cdot \bm{S}_{1j} \nonumber \\
  & = \frac{\bar{s}^2}{2} \left(t^\dagger_{l\alpha} t_{j\alpha} + t^\dagger_{l\alpha} t^\dagger_{j\alpha}  + t^\dagger_{j\alpha} t_{l\alpha}  + t_{j\alpha} t_{l\alpha} \right),
\end{align}
where the replacement $s^{(\dagger)}_i \to \bar{s} $ is introduced, and only the two-body terms of the boson operators are left.
Using the Fourier transformation and the Bogoliubov boson $\gamma^{(\dagger)}_{k\alpha}$, $\delta n^0_{k_x'} $ is expressed as
\begin{align}
  \delta n^0_{k_x'}
  &= \eta_x \bar{s}^2  \sum_p \left\{  -[\cos p + \cos (p-k_x')] \cosh \theta_{p-k_x'} \sinh \theta_p \right. \nonumber \\
  &\times   \gamma^{\dagger}_{p-k_x' \alpha}\gamma^{\dagger}_{-p \alpha}   
  +\cos p \cosh \theta_{p+k_x'} \cosh \theta_p \gamma^{\dagger}_{-p-k_x' \alpha}\gamma^{\dagger}_{p \alpha} \nonumber \\
  &+\cos (p-k_x') \sinh \theta_{p-k_x'} \sinh \theta_p \gamma^{\dagger}_{p-k_x' \alpha}\gamma^{\dagger}_{-p \alpha} \nonumber \\
  &-[\cos p + \cos (p-k_x')] \sinh \theta_{p-k_x'} \cosh \theta_p \gamma_{-p+k_x' \alpha}\gamma_{p \alpha} \nonumber \\
  &+\cos p \sinh \theta_{p+k_x'} \sinh \theta_p \gamma_{p+k_x' \alpha}\gamma_{-p \alpha} \nonumber \\
  &+ \left.
  \cos (p-k_x') \cosh \theta_{p-k_x'} \cosh \theta_p \gamma_{-p+k_x' \alpha}\gamma_{p \alpha}   
  \right\},
\end{align}
where the two-boson creation or annihilation terms are left to treat the two-triplon continuum contributing to $N^s(k'_x, k'_y=0,\omega)$.
Then $N^s(k'_x, k'_y=0,\omega)$ is represented as
\begin{align}
  & N^s(k'_x, k'_y=0,\omega) =\sum_{\alpha \ne 0} | \langle \alpha| \delta n^0_{k'_x} | 0\rangle|^2 \delta( \omega - E_\alpha + E_0) \nonumber \\
  &=6\eta_x^2 \bar{s}^4 \sum_p
  \left[ \cos \left(p + \frac{k'_x}{2} \right)  \cos \left( \frac{k'_x}{2}  \right) \sinh (\theta_p + \theta_{p+k'_x}) \right.  \nonumber \\
    & - \left. \cos p \cosh (\theta_p + \theta_{p+k'_x} )\right] \nonumber \\
  &\times \left[\cos \left(p + \frac{k'_x}{2} \right)  \cos \left( \frac{k'_x}{2}  \right) \sinh (\theta_p + \theta_{p+k'_x}) \right. \nonumber \\
    & - \left. \cos \left( p + \frac{k'_x}{2} \right)\cos \left( \frac{k'_x}{2} \right) \cosh (\theta_p + \theta_{p+k'_x} ) \right] \delta(\omega - \omega_p - \omega_{p+k'_x} ).
\end{align}

As for
\begin{align}
  \delta n^{\pi}_{k_x'}
  =  - \eta_{x} \sum_{\bm l} e^{-i k_x' l } \sum_{r = \pm 1}( \vb*{S}_{0l} \cdot \vb*{S}_{0 l+r} - \vb*{S}_{1l} \cdot \vb*{S}_{1 l+r}  ),
\end{align}
we use the same notation as
\begin{align}
  &\bm{S}_{0l} \cdot \bm{S}_{0j}  -   \bm{S}_{1l} \cdot \bm{S}_{1j} \nonumber \\
  & = - \frac{i \bar{s} }{2} \varepsilon_{\alpha\beta\delta} 
  \left(t^\dagger_{l\alpha} t^\dagger_{j\beta}t_{j\delta} + t_{l\alpha} t^\dagger_{j\beta} t_{j\delta}  + t^\dagger_{j\alpha} t^\dagger_{l\beta}t_{l\delta}  + t_{j\alpha} t^\dagger_{l\beta} t_{l\delta} \right),
\end{align}
and then obtain
\begin{align}
  \delta n^{\pi}_{k_x'}
  & =  - \frac{i \bar{s} }{2} \eta_x \varepsilon_{\alpha\beta\delta} ( 1 + e^{-ik_x'} ) \frac{1}{\sqrt{N_x}}
  \sum_{p q} \left[ e^{i(k_x'+p)} + e^{-p} \right] \nonumber \\
  & \times  (\cosh \theta_p - \sin \theta_p) ( \cosh \theta_q \sinh \theta_{k_x'+p+q} \gamma^{\dagger}_{p \alpha} \gamma^{\dagger}_{q \beta}  \gamma^{\dagger}_{-k_x' - p- q \delta}  \nonumber \\
  & + \sinh \theta_q \cosh \theta_{k_x'+p+q} \gamma_{-p \alpha} \gamma_{-q \beta}  \gamma_{k_x' + p + q \delta} ).
  \label{eq_n_pi_k}
\end{align}
It should be noted that the three-boson creation or annihilation terms are left in Eq.~(\ref{eq_n_pi_k}).
Finally, $N^s(k'_x, k'_y=\pi,\omega) $ is expressed as
\begin{align}
  &N^s(k'_x, k'_y=\pi,\omega) \nonumber \\
  &=\sum_{\alpha \ne 0} | \langle \alpha| \delta n^{\pi}_{k'_x} | 0\rangle|^2 \delta( \omega - E_n + E_n) \nonumber \\
  &=\frac{12\eta_x^2 \bar{s}^2 }{N_x} ( 1 + \cos k'_x )
  \sum_{pq} \cos \left(p + \frac{k'_x}{2} \right) \nonumber \\
  &\times \cosh \theta_q \sinh \theta_{p + q +k'_x } ( \cosh \theta_p - \sinh \theta_ p ) \nonumber \\
  &\times \left[   \cos \left(p + \frac{k'_x}{2}  \right) ( \cosh \theta_p - \sinh \theta_p ) \sinh (\theta_{p+q+k'_x+} - \theta_{q} ) \right.  \nonumber \\
    & + \cos \left(p + \frac{k'_x}{2}  \right)  ( \cosh \theta_q - \sinh \theta_q ) \sinh (\theta_{q}  - \theta_{p+q+k'_x+} )   \nonumber \\
    & + \left. \cos \left(p + q + \frac{k'_x}{2}  \right) ( \cosh \theta_{p+q+k'_x} - \sinh \theta_{p+q+k'_x}) \sinh (\theta_q - \theta_p) \right] \nonumber \\
  & \times \delta(\omega - \omega_p - \omega_q - \omega_{p+q+k'_x} ).
\end{align}

%----------------------------------

\end{document}